\documentclass{pasj00}
\draft

\begin{document}
\SetRunningHead{M. Kawada et al.}{The Far-Infrared Surveyor (FIS)}
\Received{2007/**/**}
\Accepted{2007/**/**}

\title{The Far-Infrared Surveyor (FIS) for AKARI\thanks{AKARI is a 
JAXA project with the participation of ESA}}

\author{Mitsunobu \textsc{Kawada},\altaffilmark{1}
	Hajime \textsc{Baba},\altaffilmark{2}
	Peter D. \textsc{Barthel},\altaffilmark{3}
	David \textsc{Clements},\altaffilmark{4}
	Martin \textsc{Cohen},\altaffilmark{5}
	Yasuo \textsc{Doi},\altaffilmark{6}
	Elysandra \textsc{Figueredo},\altaffilmark{7}
	Mikio \textsc{Fujiwara},\altaffilmark{8}
	Tomotsugu \textsc{Goto},\altaffilmark{2}
	Sunao \textsc{Hasegawa},\altaffilmark{2}
	Yasunori \textsc{Hibi},\altaffilmark{9}
	Takanori \textsc{Hirao},\altaffilmark{1}
		\thanks{Present Address is Japan Science and Technology Agency, 
			4-1-8 Honcho, Kawaguchi, Saitama 332-0012, Japan}
	Norihisa \textsc{Hiromoto},\altaffilmark{10}
	Woong-Seob \textsc{Jeong},\altaffilmark{2}
	Hidehiro \textsc{Kaneda},\altaffilmark{2}
	Toshihide \textsc{Kawai},\altaffilmark{11}
	Akiko \textsc{Kawamura},\altaffilmark{1}
	Do \textsc{Kester},\altaffilmark{12}
	Tsuneo \textsc{Kii},\altaffilmark{2}
	Hisato \textsc{Kobayashi},\altaffilmark{2,13}
	Suk Minn \textsc{Kwon},\altaffilmark{14}
	Hyung Mok \textsc{Lee},\altaffilmark{15}
	Sin'itirou \textsc{Makiuti},\altaffilmark{2}
	Hiroshi \textsc{Matsuo},\altaffilmark{9}
	Shuji \textsc{Matsuura},\altaffilmark{2}
	Thomas G. \textsc{M\"{u}ller},\altaffilmark{16}
	Noriko \textsc{Murakami},\altaffilmark{1}
	Hirohisa \textsc{Nagata},\altaffilmark{2}
	Takao \textsc{Nakagawa},\altaffilmark{2}
	Masanao \textsc{Narita},\altaffilmark{2}
	Manabu \textsc{Noda},\altaffilmark{17}
	Sang Hoon \textsc{Oh},\altaffilmark{15}
	Yoko \textsc{Okada},\altaffilmark{2}
	Haruyuki \textsc{Okuda},\altaffilmark{2}
	Sebastian \textsc{Oliver},\altaffilmark{18}
	Takafumi \textsc{Ootsubo},\altaffilmark{1}
	Soojong \textsc{Pak},\altaffilmark{19}
	Yong-Sun \textsc{Park},\altaffilmark{15}
	Chris P. \textsc{Pearson},\altaffilmark{2,20}
	Michael \textsc{Rowan-Robinson},\altaffilmark{4}
	Toshinobu \textsc{Saito},\altaffilmark{2,13}
	Alberto \textsc{Salama},\altaffilmark{20}
	Shinji \textsc{Sato},\altaffilmark{1}
	Richard S. \textsc{Savage},\altaffilmark{18}
	Stephen \textsc{Serjeant},\altaffilmark{7}
	Hiroshi \textsc{Shibai},\altaffilmark{1}
	Mai \textsc{Shirahata},\altaffilmark{2}
	Jungjoo \textsc{Sohn},\altaffilmark{15}
	Toyoaki \textsc{Suzuki},\altaffilmark{2,13}
	Toshinobu \textsc{Takagi},\altaffilmark{2}
	Hidenori \textsc{Takahashi},\altaffilmark{21}
	Matthew \textsc{Thomson},\altaffilmark{18}
	Fumihiko \textsc{Usui},\altaffilmark{2}
	Eva \textsc{Verdugo},\altaffilmark{20}
	Toyoki \textsc{Watabe},\altaffilmark{11}
	Glenn J. \textsc{White},\altaffilmark{7,22}
	Lingyu \textsc{Wang},\altaffilmark{4}
	Issei \textsc{Yamamura},\altaffilmark{2}
	Chisato \textsc{Yamamuchi},\altaffilmark{2}
	and 
	Akiko \textsc{Yasuda}\altaffilmark{2,23}}

\email{kawada@u.phys.nagoya-u.ac.jp}

\altaffiltext{1}{Graduate School of Sciences, Nagoya University,
	Furo-cho, Chikusa-ku, Nagoya 464-8602, Japan}
\altaffiltext{2}{Institute of Space and Astronautical Science, JAXA,
	3-1-1 Yoshinodai, Sagamihara 229-8510, Japan}
\altaffiltext{3}{Groningen Kapteyn Institute, Rijksuniversiteit, 
	Landleven 12, Postbus 800, 9700 AV Groningen, Netherlands}
\altaffiltext{4}{Imperial College, London, Blackett Laboratory,
	Prince Consort Road, London, SW7 2AZ, UK}
\altaffiltext{5}{Radio Astronomy Laboratory, 601 Campbell Hall,
	University of California, Berkeley, CA94720, USA}
\altaffiltext{6}{Department of General System Studies, 
	Graduate School of Arts and Sciences, 
	The University of Tokyo, 3-8-1 Komaba, Meguro-ku, 
	Tokyo 153-8902, Japan}
\altaffiltext{7}{Department of Physics and Astronomy, Faculty of Science,
	The Open University, Walton Hall, Milton Keynes, MK7 6AA, UK}
\altaffiltext{8}{Advanced Communications Technology Group,
	New Generation Network Research Center,
	NICT, 4-2-1 Nukui-Kitamachi, Koganei, Tokyo 184-8795, Japan}
\altaffiltext{9}{Advanced Technology Center, 
	National Astronomical Observatory of Japan,
	2-21-1 Osawa, Mitaka, Tokyo 181-8588, Japan}
\altaffiltext{10}{Optelectronics and Electromagnetic Wave Engineering, 
	Shizuoka University, 3-5-1 Johoku, Hamamatsu, Japan}
\altaffiltext{11}{Technical Center of Nagoya University, Furo-cho, Chikusa-ku,
	Nagoya 464-8601, Japan}
\altaffiltext{12}{Netherlands Institute for Space Reasearch SRON, 
	Landleven 12, PO Box 800, 9700 AV Groningen, Netherlands}
\altaffiltext{13}{Department of Physics, Graduate School of Science, 
	The University of Tokyo, Bunkyo-ku, Tokyo 113-0033, Japan}
\altaffiltext{14}{Department of Science Education, 
	Kangwon National University, 
	192-1 Hyoja-Dong, Chuncheon, Gangwon-Do, 200-701, Korea}
\altaffiltext{15}{Astronomy Program, Department of Physics and Astronomy, 
	FPRD, Seoul National University,
	Shillim-dong, Kwanak-gu, Seoul 151-742, Korea}
\altaffiltext{16}{Max-Planck-Institut f\"{u}r extraterrestrische Physik, 
	Giessenbachstrasse, 85748 Garching, Germany}
\altaffiltext{17}{Nagoya City Science Museum, 2-17-1 Sakae, Naka-ku, 
	Nagoya 460-0008, Japan}
\altaffiltext{18}{Astronomy Centre, University of Sussex, Falmer, 
	Brighton, BN1 9QJ, UK}
\altaffiltext{19}{Department of Astronomy and Space Science, 
	Kyung Hee University, 
	Yongin-si, Gyeonggi-do 446-701, Korea}
\altaffiltext{20}{European Space Astronomy Centre, ESA, 
	Villanueva de la Canada, Post Box 78, 28691 Madrid, Spain}
\altaffiltext{21}{Gunma Astronomical Observatory, 6860-86 Nakayama, 
	Takayama-mura, Agatsuma-gun, Gunma 377-0702, Japan}
\altaffiltext{22}{Space Science and Technology Division, The Rutherford 
	Appleton Laboratory, Chilton, Didcot, Oxfordshire OX11 0QX, England}
\altaffiltext{23}{The Graduate University for Advanced Studies, 
	Shonan Village, Hayama, Kanagawa 240-0193, Japan}



\maketitle

\begin{abstract}
The Far-Infrared Surveyor (FIS) is one of two focal plane instruments on the 
AKARI satellite.  FIS has four photometric bands at 65, 90, 140, and 160 $\micron$, 
and uses two kinds of array detectors.  The FIS arrays and optics are designed to 
sweep the sky with high spatial resolution and redundancy. The actual scan width 
is more than eight arcmin, and the pixel pitch is matches the diffraction 
limit of the telescope.  Derived point spread functions (PSFs) from observations 
of asteroids are similar to the optical model. Significant excesses, however, are 
clearly seen around tails of the PSFs, whose contributions are about 30\% of the 
total power.  All FIS functions are operating well in orbit, and its performance 
meets the laboratory characterizations, except for the two longer wavelength bands, 
which are not performing as well as characterized.  Furthermore, the FIS has 
a spectroscopic capability using a Fourier transform spectrometer (FTS).  
Because the FTS takes advantage of the optics and detectors of the photometer, it can 
simultaneously make a spectral map.  This paper summarizes the in-flight technical 
and operational performance of the FIS.
\end{abstract}

\section{Introduction}

The first extensive survey of the far-infrared sky was made by the {\it 
Infrared Astronomy Satellite} (IRAS) launched in 1983, more than two decades 
ago.  IRAS provided point source catalogs, as well as infrared sky maps, for 
almost the entire sky, revealing the infrared view of the universe.   The IRAS 
products became a standard dataset, not only for infrared astronomy but also for 
many other fields. 
With the widespread use of the IRAS dataset, observations in the infrared are now 
considered an irreplaceable tool. Compared with the datasets at other 
wavelengths, the IRAS dataset appears rather shallow, although it provides an 
unbiased and wide area coverage survey.  However, a next-generation infrared 
dataset is needed to push the frontier of astrophysics.  The Far-Infrared 
Surveyor (FIS) on the AKARI satellite \citep{Shibai07, Murakami07} was 
developed to provide this new dataset in the far-infrared region, taking 
advantage of recent technology, including the cryogenics \citep{Nakagawa07}.
The FIS was designed to perform an all-sky survey in the far-infrared region 
with higher spatial resolution and higher sensitivity than the IRAS.  A 
combination of the Infrared Camera (IRC) \citep{Onaka07} on the AKARI 
satellite also enables a wider wavelength coverage than IRAS.

The {\it Spitzer Space Telescope} (SST) \citep{Werner04}, which is used 
extensively in the observation of various objects in the infrared region, has 
brought new insights into the universe with its high spatial resolution and high 
sensitivity.  FIS is a complementary instrument because of its capability of 
covering wide areas, and its multiple photometric bands.  Furthermore, the FIS has 
the advantage of allowing far-infrared spectroscopy with a Fourier transform 
spectrometer (FTS).  The {\it Infrared Space Observatory} (ISO) \citep{Kessler96} 
demonstrated the potential of spectroscopy in the infrared region for 
diagnostics of the interstellar medium and radiation field.  Adopting 
two-dimensional array detectors, the FTS of FIS works as an imaging FTS. 
Consequently, it affords high efficiency observations of the spatial structure 
in spectra. 

This paper describes the design and operation of FIS in two sections, and then 
discusses its flight performance and advantage.

\section{Instrument Design}
\begin{figure}
  \begin{center}
    \FigureFile(80mm,70mm){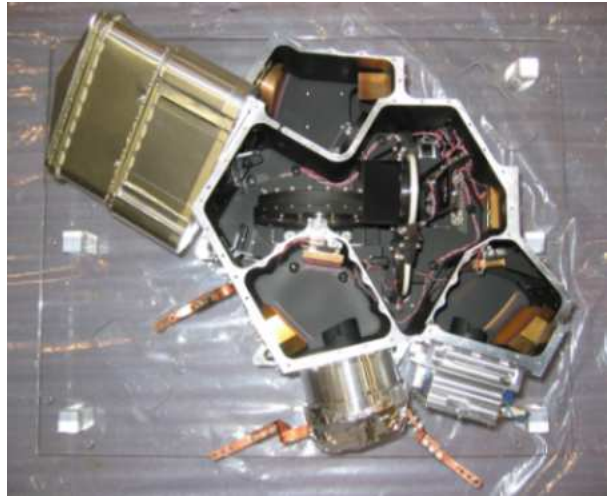}
  \end{center}
  \caption{A picture of FIS at final integration. The top cover is removed 
	to show the FIS optics. The orientation of the FIS is the same as 
	in Fig.\ref{fig:FIS_optics}}
  \label{fig:FIS}
\end{figure}

The FIS is a composite instrument, consisting of a scanner and a spectrometer. 
Adopting the newly developed large format array detectors, FIS achieves high 
spatial resolution and sensitivity. Fig.\ref{fig:FIS} shows a picture of FIS 
during final integration; the top cover is removed to show the interior.

\subsection{Optical Design}
\begin{figure*}
  \begin{center}
    \FigureFile(150mm,100mm){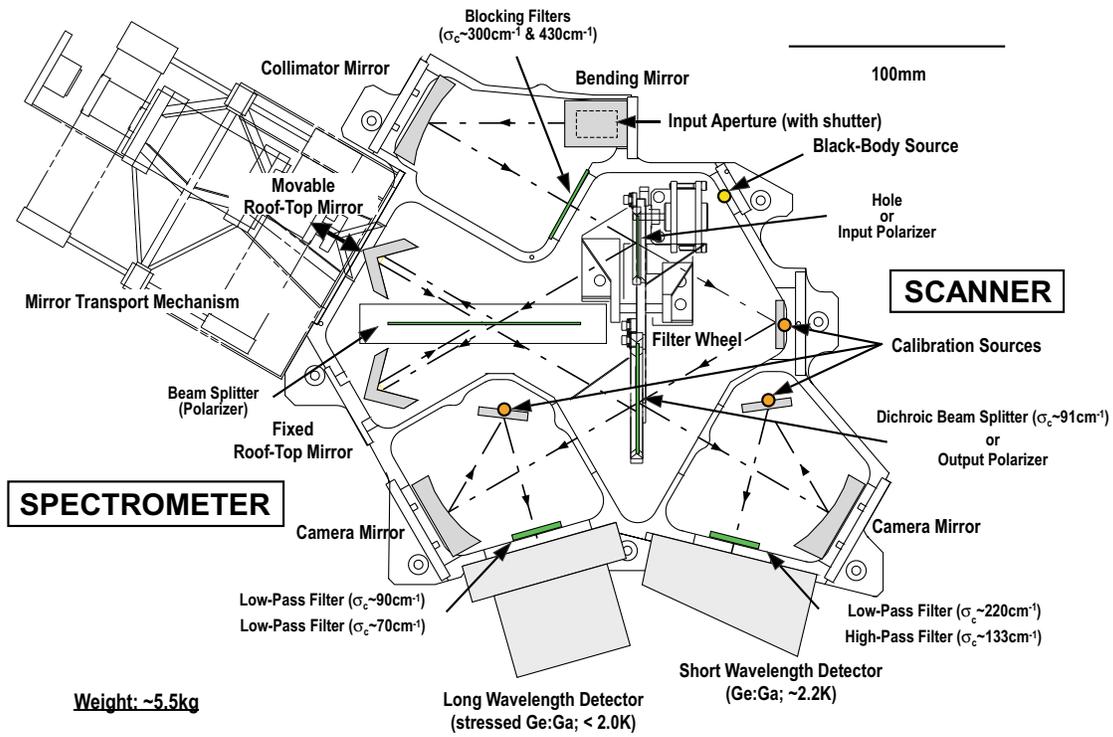}
  \end{center}
  \caption{Drawings of the FIS optical configuration. The major optical 
	components are shown with labels. The FIS instrument provides two 
	functions --- the scanner (left side) and the spectrometer (right 
	side), which use different optical paths.  Each function is 
	selected by rotating the filter wheel. }
  \label{fig:FIS_optics}
\end{figure*}

Fig.\ref{fig:FIS_optics} illustrates the FIS optical design.  FIS is about 
50 cm along the major axis and weighs about 5.5 kg.  To reduce the size 
and resources required, the scanner and spectrometer share some optical 
components and detector units.  Rotating the filter wheel selects the 
appropriate function.

Light coming from the telescope is focused near the FIS input aperture.  
After passing through the input aperture, the beam is bent into the
FIS optical plane and led to the collimator mirror. The collimated beam goes 
to the filter wheel, which selects the scanner or the spectrometer by 
choosing the filter combination.  In the scanner mode, it selects a the 
combination of an open hole and a dichroic beam splitter.  The beam passes 
through the hole, and is reflected by a flat mirror to the dichroic filter 
on the filter wheel. Longer wavelength photons ($>110$ \micron) pass through 
the filter and shorter wavelength photons are reflected.  Camera optics 
focus the beams onto the detector units according to their wavelength.

A Fourier transform spectrometer (FTS) serves as the spectroscopic component. 
A polarizing Michelson interferometer, the so-called 
Martin--Puplett interferometer \citep{Martin69}, is employed in the FTS 
optics.  This type of interferometer requires input and output polarizers 
to provide linear polarization. These two polarizers are on the filter 
wheel, and are selected in the spectrometer mode. The collimated beam is 
reflected by the input polarizer to the interferometer, while the remaining 
component is absorbed by a blocking wall. A beam splitter divides the 
linearly polarized beam into two beams. The beam splitter is also a 
polarizer, whose polarizing angle is rotated by 45 degrees relative to the 
incident linear polarized beam. All polarizers are wire-grid filters printed 
on thin Mylar films supplied by QMC Instruments Ltd. The divided beams are 
reflected by roof-top mirrors; one is fixed and the other is
movable to change the optical path difference between two beams. After that, 
the beam splitter recombines the two beams, and the interfered beam goes 
to the output polarizer. Finally, the elliptically polarized beam is 
separated into two axis components by the output polarizer. Each component 
is focused on the corresponding detector unit.

FIS has a cold shutter at the input aperture to allow measurement of the 
dark current of the detector. For the calibration of the detectors, there are 
three light sources in the FIS housing. One is in the light path of the 
scanner, and the others irradiate the front of each detector.  The 
intensity and emitting duration of each lamp are independently controlled by 
commands independently. These lamps can also simulate the light curve of the point 
source in the all-sky survey.  Since the simulated light curve is quite 
stable and reproducible, the detector responses to a point source can be 
monitored at any time. In addition to these calibration lamps, there is a 
blackbody source, whose temperature is controllable up to 40K, opposite 
the interferometer input that is used to check that operation of the FTS. 
Furthermore, 
to improve the transient response of the long wavelength detector, a 
background light source, with controllable power, is placed on the detector 
unit to continuously irradiate detector pixels.  At the current setting, 
the incident power corresponds to a sky brightness of $\sim$100 MJy/sr.

\subsection{Instrumentation Description}

\subsubsection{Field-of-View}

We use two types of photoconductive detector array to cover the far-infrared 
wavelengths (50 -- 180 $\micron$).  One is the direct-hybrid monolithic Ge:Ga 
array \citep{Fujiwara03} for the shorter wavelength of 50 to 110 $\micron$ 
(labeled SW), which was developed by the National Institute of Information 
and Communications Technology (NICT).  The Ge:Ga monolithic array is 
bump-bonded by Indium with Silicon-based cryogenic readout electronics 
\citep{Nagata04}.  The other is a compact stressed Ge:Ga array for longer 
wavelengths of 110 to 180 $\micron$  (labeled LW), which is an evolved 
version of a previous model \citep{Doi02}. The LW detector also employs 
cryogenic readout electronics. All Ge:Ga chips were supplied by NICT.

\begin{figure}
  \begin{center}
    \FigureFile(80mm,105mm){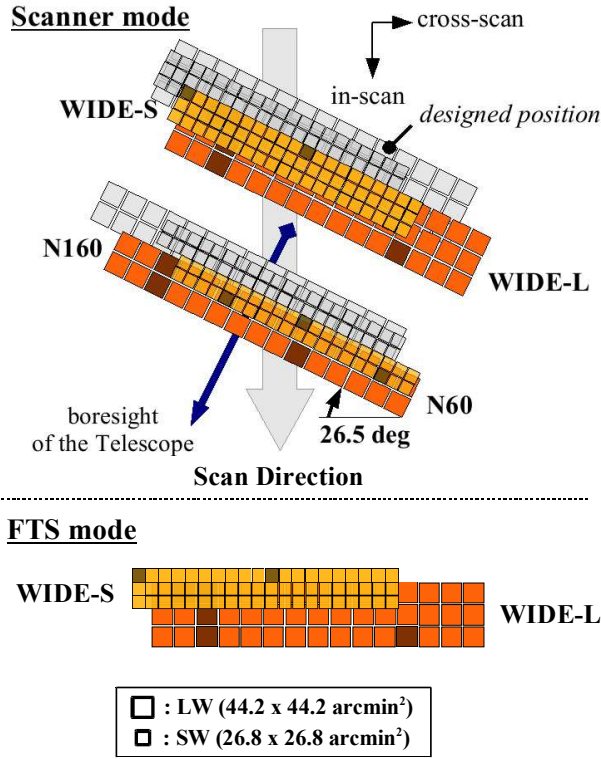}
  \end{center}
  \caption{Details of the FOVs of FIS. The upper panel shows the FOVs in 
	the scanner mode. The FOVs measured in orbit are drawn on the 
	designed FOVs with light gray. The detector's field-of-views are 
	shifted from the designed location, which is referred to the 
	boresight of the telescope. The misalignments of each array are 
	about one pixel.  The major axis of the arrays is rotated by 26.5 
	degrees in the scan direction to improve the grid pitch in the 
	cross-scan direction.  The lower panel shows FOVs in the 
	spectrometer mode, in which WIDE-S and WIDE-L are active.  The 
	misalignment of each array is larger than in the scanner mode. 
	The accuracy of the measured alignment is about half of a pixel 
	size. Inactive pixels are indicated by hatching. }
  \label{fig:array_format}
\end{figure}

There are two arrays in each detector unit for the different photometric 
bands --- WIDE-S and N60 for the SW detector and WIDE-L and N160 for the LW 
detector.  The array formats are $3 \times 20$, $2 \times 20$, $3 \times 15$, 
and $2 \times 15$ for WIDE-S, N60, WIDE-L, and N160, respectively, as shown 
in Fig.\ref{fig:array_format}.  Several pixels did not work properly after 
fabrication (shown by hatching in the figure); however, no additional bad 
pixels appeared after the launch.

The fields-of-view (FOVs) of WIDE-S and WIDE-L, N60 and N160 detectors 
overlap, although their coverage differs.  The pixel scales are designed to 
be comparable to the telescope's diffraction limits.  The arrays are 
rotated by 26.5 degrees relative to the scanning direction.  Because of 
this configuration, the width of the stripe swept out on the sky is reduced to 
about $90 \%$ of the array width, although the spatial sampling grid 
becomes half of the pixel pitch in the cross-scan direction.

The FOVs of the SW and LW detectors were measured in space by observing point 
sources.  The measured FOVs are shifted from the designed 
position by about 1 arcmin in both in-scan and cross-scan directions 
referred to the telescope's boresight (see top panel of 
Fig.\ref{fig:array_format}).  In the spectrometer mode (bottom panel of 
Fig.\ref{fig:array_format}), the misalignment of each array is larger than 
in the scanner mode, which limits the observational efficiency of spectral 
mapping.  

The distortion of the FOVs due to the FIS optics is evaluated by optical 
simulation. Although the actual distortion and magnification factor of the 
FIS optics do not match precisely, no significant discrepancy with the 
design is indicated.

\subsubsection{System Spectral Response}

The four photometric bands of FIS are defined by the combination of the 
optical filters and the spectral response of the detectors, as the 
incident photons reach the detectors through the optical filters.  The 
collimated beam passes through two blocking filters that block 
the mid- and near-infrared photons contributed mainly by stars. The 
dichroic filter then divides the beam in the frequency domain: higher 
frequency photons ($> 91$ cm$^{-1}$ in wavenumber) are reflected and 
lower frequency photons are transmitted. Finally, two filters on the 
front of each detector shape the photometric bands. Four photometric 
bands cover the 50 to 180 $\micron$ wavelength region; two are wide 
bands (WIDE-S and WIDE-L) and the other two are narrow (N60 and N160). 

In the spectrometer mode, the dichroic beam splitter is replaced by a 
combination of three polarizers. The other optical filters are the same 
as for the scanner mode. Only the wide bands (WIDE-S and WIDE-L) are used 
for spectroscopy. The spectrometer could in principle take interferograms with 
narrow bands, although some of the outer pixels of the arrays will vignette 
the telescope beam.

\begin{figure}
  \begin{center}
   \FigureFile(80mm,55mm){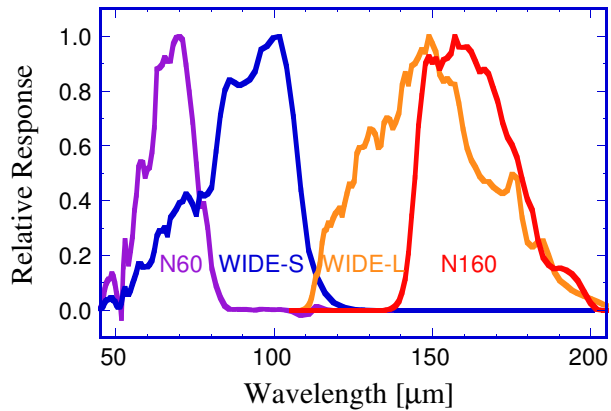}
  \end{center}
   \caption{ Plots of the system spectral responses of the FIS photometric 
	bands.  These shapes are obtained from spectral measurements of 
	optical components and detectors. The achieved optical efficiency 
	is nearly 50\% for all bands, except N60.  This plot is typical, 
	however, pixel-to-pixel variation can be seen.}
  \label{fig:FIS_bands} 
\end{figure}

Fig.\ref{fig:FIS_bands} shows the FIS system's spectral response in 
the photometric mode.  The responses are normalized at the peak.  The 
filters and optics were measured in an end-to-end system configuration 
at room temperature, although the narrow band filters for N60 and N160 were 
measured individually at cryogenic temperature, since their properties 
depend on temperature.

The spectral responses of the detectors were evaluated by a 
spectrometer, each pixel of the detector array having a different spectral 
response. This difference is larger in the LW detector, due to 
non-uniformity of the effective stress on the Ge:Ga chips.  The plot in 
Fig.\ref{fig:FIS_bands} provides typical profiles.  The spectral 
response of each pixel must be used to calculate precise pixel-to-pixel 
color corrections. 

The spectral response can also be measured by the FIS spectrometer itself, 
using the internal and external blackbody sources at different 
temperatures.  In orbit, we confirmed the system spectral response of FIS 
in this manner. The spectra of the internal 
blackbody source taken in orbit are within $10\%$ of those measured in 
the laboratory after scaling the responsivity. Since the filters and 
optics are considered stable, the system spectral response of each 
photometric band is expected to be similar to that measured in the 
laboratory as shown in Fig.\ref{fig:FIS_bands}.

The blocking filters are expected to avoid leakage of the mid- and 
near-infrared photons. The blocking efficiency required from the 
scientific observations is  $10^{-5}$ at 10 $\micron$  and $10^{-9}$ at 
0.5 $\micron$, which are realized optimally, and will in future 
be checked by observations of well-known stars.

\section{Observation Modes}

\subsection{All-Sky Survey}

FIS is designed primarily to perform an all-sky survey with four 
photometric bands. The goal is to observe the entire sky for at least two 
independent orbits.  The first half year following the performance 
verification phase is dedicated to observations for the all-sky survey. 
During the remaining life, supplemental survey observations will fill in 
the incomplete sky areas, sharing the observation time with dedicated 
science programs.

During the survey, the detectors are read out continuously with a constant 
sampling rate for each array, corresponding to about three samples in a 
pixel crossing time.  Detectors are reset to discharge the photo current 
at appropriate intervals of about 2 sec nominally, 0.5 sec for bright 
sky, and for each 
sampling (correlated double sampling: CDS) at the Galactic plane, whose 
reset intervals correspond to about 26 and 45 ms for the SW and LW 
detectors, respectively.  Calibration flashes with illuminator lamps are 
inserted periodically every minute while keeping the shutter open to trace 
the detector responsivity. Near the ecliptic poles, where the detectors 
sweep frequently, a one-minute calibration sequence with the shutter 
closed is executed for nearly every orbit to monitor the long term trend 
of the detector responsivity.

\subsection{Pointing Observation}

Before launch of the AKARI satellite, the project science team members 
selected the core programs and target sources were selected. In addition 
to the core 
programs, some portion of the observation time was open to the community.
For the scientific programs and the Open Time proposals, the AKARI 
instruments operate in a pointing mode.  The FIS observations are 
categorized in three major astronomical observation templates (AOTs), 
--- two for photometry and one for spectroscopy.  Details of each AOT are 
described below.

\subsubsection{FIS01: Compact Source Photometry}

The FIS01 template is used to observe point-like or small scale sources. In 
this AOT, for one pointing observation, the detectors sweep the sky two 
times in round trips.  Between two round trips, the scan path is shifted by 
either a few pixels (70 arcsec) or half of the FOV (240 arcsec) in the cross-scan 
direction, which is selectable.  The other adjustable parameters are the 
reset interval and scan speed.  The scan speed is selectable from 8"/sec or 
15"/sec, which are nearly 14 to 30 times slower than that of the all-sky 
survey.  The detection limits should be improved by a factor of the 
exposure time or more using the charge integration amplifier.  Furthermore, 
using a slower scanning speed reduces the transient response effects of the 
detector response.

\begin{figure*}
  \begin{center}
   \FigureFile(160mm,100mm){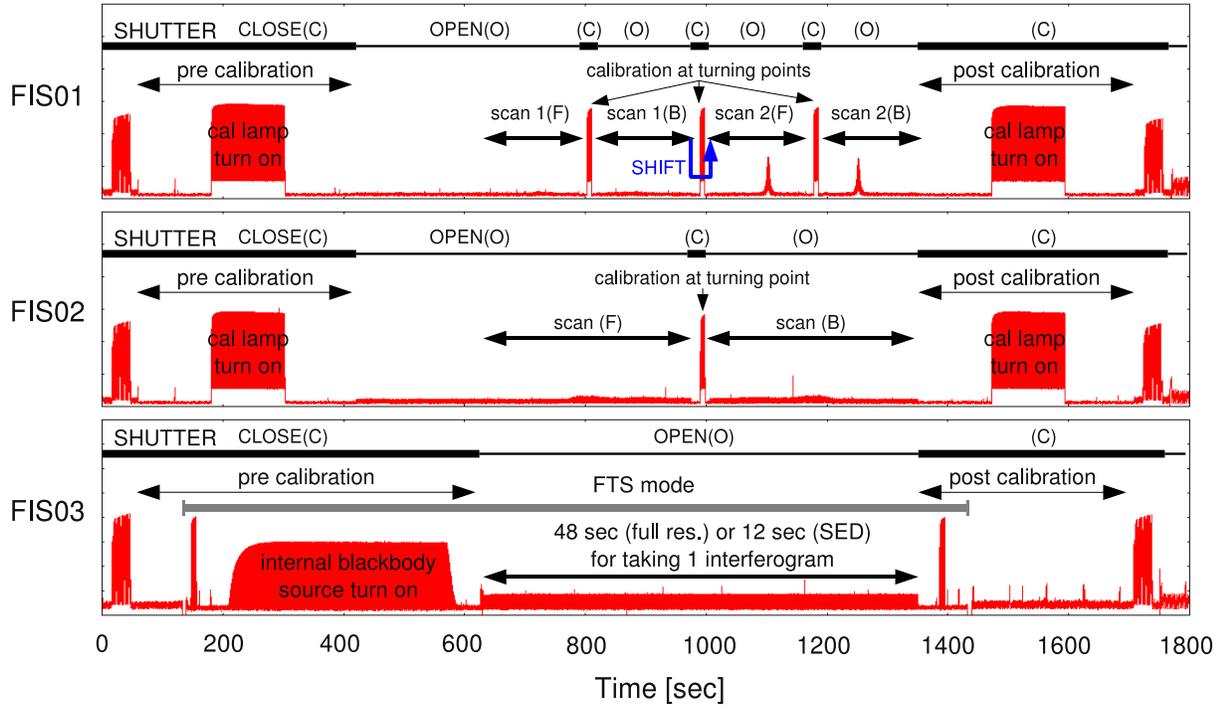}
  \end{center}
   \caption{Observation sequences of each AOT are illustrated on the real 
	signal of the SW detector. The top panel shows the sequence for 
	FIS01, which has two round trips with different offsets to the 
	target position in the cross-scan direction.  The middle panel is 
	for FIS02, which has one round trip. In pre- and post-calibration, 
	calibration lamps are turned on for two minutes with the shutter 
	closed. At the turning points, calibration lamps are illuminated 
	for about eight seconds with the shutter closed.  The bottom panel 
	indicate the sequence of FIS03, which is for spectroscopy. During 
	pre calibration, the internal blackbody source is activated, and 
	the reference spectra are taken for about six minutes. In the 
	following 12 minutes, sky spectra are observed continuously. 
	Signals before pre-calibration and after post-calibration are the 
	one-minute calibration for the all-sky survey.
   }
  \label{fig:FIS_AOT} 
\end{figure*}

An observation sequence takes about 30 minutes as shown in the top panel 
of Fig.\ref{fig:FIS_AOT}.  During about 10 minutes of maneuvering to the 
spacecraft to point toward the target, the standard calibration sequence 
is executed with the 
shutter closed, i.e., measuring dark current, illuminating the calibration 
lamps continuously for about two min and flashing the calibration lamps 
several times.  After opening the shutter and waiting 210 sec, the 
scanning sequence begins and continues for about 12 minutes.  Then, the 
shutter is closed again and the maneuver to the all-sky survey begins 
with the post observation calibrations.  During scanning observations, at 
the scan turning points, the shutter is closed for 30 sec and the 
calibration lamps are turned on for about eight sec to monitor the drift 
of the detector responsivity during the pointing observation.
 
\subsubsection{FIS02: Wide Area Mapping}

FIS02 is the template for wide area mapping. This AOT executes only one 
round trip.  The detectors sweep a longer strip than for FIS01, but the 
detection redundancy is reduced. Overlapping scans are critical for high 
quality wide area mapping.  By selecting the 15"/sec scan speed, the 
strip length reaches over one degree.  The calibration sequences in the 
pre- and post observation phase and at the turning point are the same as 
for FIS01.  The middle panel of Fig.\ref{fig:FIS_AOT} 
illustrates the observation sequence.

\subsubsection{FIS03: Imaging Spectroscopy}

FIS03 is the template for spectroscopic observations.  The bottom panel of 
Fig.\ref{fig:FIS_AOT} illustrates the observation sequence. In this AOT, 
the target is locked on the detectors. To use the FTS, the optics are
switched to FTS mode by rotating the filter wheel during the maneuver to 
the target position.  The sampling sequence changes for the FTS and the 
movable mirror starts to operate.  Important parameters in FIS03 are the 
spectral resolution and array selection.  Users choose from two spectral 
resolution modes: a full resolution mode and a low resolution mode (named 
SED mode), with spectral resolution of 0.19 cm$^{-1}$ and 1.2 cm$^{-1}$ 
without apodization, respectively.  Taking one interferogram in the full 
resolution mode takes four times longer than that in the SED mode.  
Consequently, in one pointing observation, 15 full resolution or 59 SED 
mode interferograms can be taken.  The other important parameter is the 
array selection, which was added after launch, because the FOVs of the two 
detectors are misaligned as shown in Fig.\ref{fig:array_format}. Depending 
on the position of the target on the detectors, there are three choices: 
nominal, SW, and LW positions.

During the maneuver to the target position, the internal blackbody source 
is turned on at the proper temperature, with the shutter closed, and 
interferograms are taken as reference spectra.  A short calibration 
sequence using the calibration lamps is also conducted at the pre- and 
post observation phase.  After the observation, the observation mode 
changes to photometry mode for the all-sky survey, during the recovery 
maneuver to the all-sky survey.

\bigskip
Parameters for each AOT are summarized in Table \ref{tab:aot_list}.

\subsubsection{Parallel Observation}

FIS can operate in parallel with the IRC observations. Since the FOVs of FIS 
are separated from the FOVs of the IRC by about half a degree, they can not 
observe the same target.  Nevertheless, taking data with FIS is 
interesting in many cases as a serendipitous survey.  The nominal 
operation of the IRC is to make observations with long exposure times to 
provide deep images or spectra.  Since the FIS photometry is designed for 
scanning mode, the pointed observations of the photometry are ineffective 
due to the narrow array formats and larger pixel scales.  Therefore, FIS 
is operated in the spectrometer mode, if the sky is bright enough to 
detect signals by the FTS. The observation sequence for parallel 
observation is the same as for FIS03.  These parallel observations are 
productive, especially for the Large Magellanic Cloud, where IRC is making 
systematic surveys, as well as toward bright complex regions like the 
Galactic plane.

Specific calibration sequences operates during the remaining IRC oriented 
observations.  Once a day, the calibration sequence to evaluate 
detector transient response is executed.  About once every three days, 
the calibration lamp stability is also measured.  These calibration data 
are used to track long term trends in detector performance.

\section{Flight Performance}

In orbit, all FIS functions are working as designed. The cold shutter and 
the filter wheel have operated constantly while in orbit, more than ten 
thousand times and a few hundred times, respectively.  All the functions 
are controlled by the onboard electronics without any difficulty.  In the 
following subsections, the FIS flight performance is described.

\subsection{Imaging Quality}

\begin{figure*}
  \begin{center}
   \FigureFile(150mm,100mm){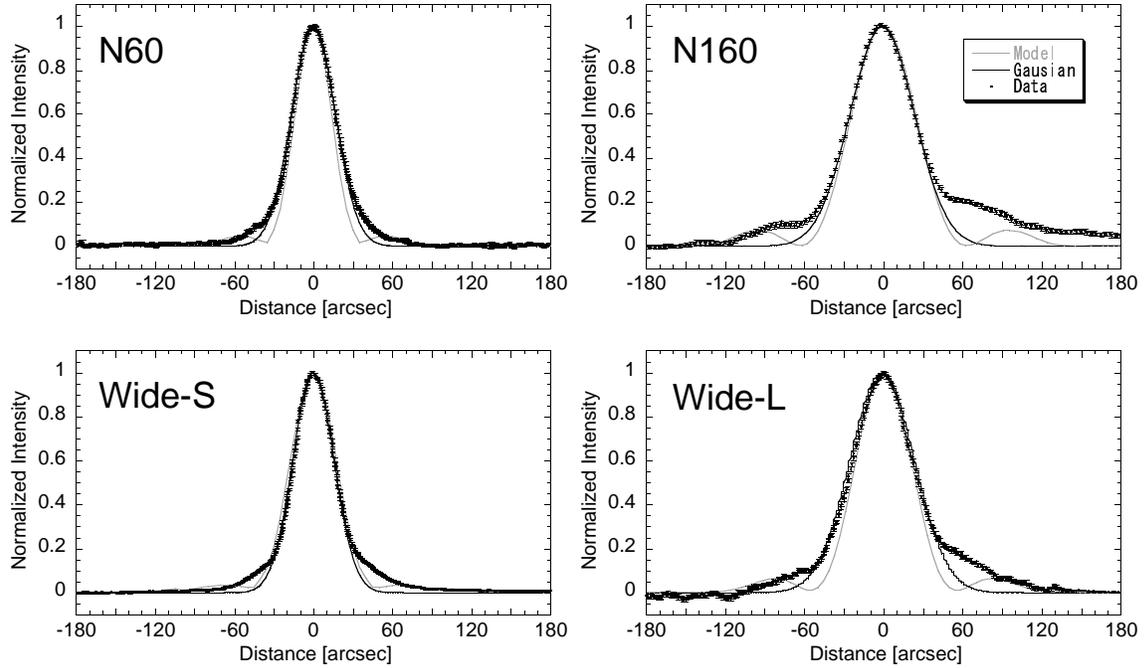}
  \end{center}
   \caption{The PSF for each band constructed from the observations of 
	asteroids (data points) with the Gaussian fit (thin line) at the 
	central part. Expected PSFs from the optical model are indicated 
	by dotted lines.  There are significant enhancement at the tails 
	of the PSFs in all bands.
   }
  \label{fig:FIS_psf} 
\end{figure*}

The point spread functions (PSFs) of FIS were measured in the laboratory 
using a pin hole source. Widths of the measured PSFs were almost consistent 
with those expected from the optical simulation. In orbit, the system PSFs 
including the telescope system are constructed from observations of bright 
point sources.  The PSFs are similar to the laboratory measurements.  
As shown in Fig.\ref{fig:FIS_psf}, the PSFs conform to the estimations 
from the optical model at more than the half maximum of the peak.  The 
full widths at the half maximum of the PSFs, derived from Gaussian 
fitting, are summarized in Table \ref{tab:performance}. At the tails of the 
PSFs, there are significant enhancements, whose power is about 30\% of the 
total power.  This extended halo is a cause for the degradation of source 
detection.

In the spectrometer mode, the PSFs are evaluated by slow scan observations of 
the point sources with the spectrometer optics, and are about $20\%$ wider 
than in the photometer mode. It is, however, possible to take spectra for 
each pixel and images with nearly one arc minute spatial resolution, 
simultaneously.

\begin{figure*}
  \begin{center}
   \FigureFile(150mm,70mm){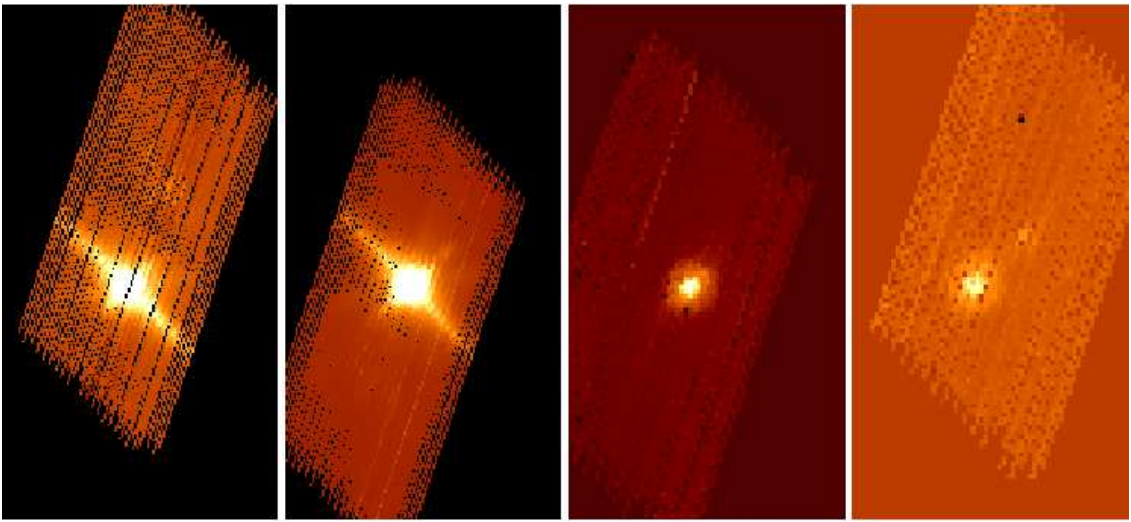}
  \end{center}
   \caption{A demonstration of the imaging quality of FIS.  The raw image 
	of the asteroid {\it Ceres} for each band is shown in each panel, 
	observed by FIS01. The panels are N60, WIDE-S, WIDE-L, and N160 
	(from left to right).  The color scale is modified to enhance the 
	lower signal level. The ghost signals are seen in all bands, 
	especially for narrow bands. In the images of the SW detector, 
	cross talk signal is clearly seen along both array axes.
   }
  \label{fig:FIS_ghost} 
\end{figure*}

An additional factor degrades the imaging quality of the SW detector, 
namely, considerable cross talk between pixels in both axes of the 
arrays.  
One possibility to explain this phenomenon is that the incident 
far-infrared photons diffuse into the monolithic Ge:Ga array 
during multiple reflection on the front and back surfaces of the 
detector substrate.
Furthermore, both 
the SW and the LW detectors show significant ghost signals.  The 
reason for the ghost is, presumably, electrical cross talk in the 
multiplexer of the cryogenic readout electronics. In this case, the ghost 
appears in other arrays of the same detector.  The images of the asteroid 
{\it Ceres} observed by FIS01 are shown in Fig.\ref{fig:FIS_ghost}, as an 
example. To enhance the effect of the cross talk, the color level has been 
adjusted.  Since the position and strength of the cross talk are stable, 
it should be possible to remove it from the original.

Another possible degradation of the image quality comes from the 
detector's transient response. For pointed observations, the detector 
scans a source, both forward and backward on the same pixel.  The effect 
of transient response has been evaluated for each observation, and is not 
significant for slow scan observations.

\subsection{Detector Performance}

The readout method for the FIS detectors is based on a Capacitive 
Trans-Impedance Amplifier (CTIA) using newly developed cryogenic devices 
\citep{Nagata04}.  Although the CTIA has a wide dynamic range, the 
linearity is rather poor, and the effective bias on the detectors drifts. In 
laboratory measurements, the relation between the output signal and the 
amount of the stored charge was well calibrated, and the relation is 
confirmed in orbit.  Therefore, the photo current can be accurately 
reproduced from the output signal.

Charged particle hits are another important influence on detector 
performance.  About once a minute, charged particles hit a pixel, and in 
some cases, the responsivity drifts for between several seconds and several 
minutes.  Near the South Atlantic Anomaly (SAA), the hit rate of charged 
particles is too high to observe the sky signal. After passing the SAA 
region, the detector responsivity increases significantly and relaxes 
gradually with a decay on the order of hours.  To cure the effect of the 
SAA, bias boosting is applied just after passing through the SAA.  By 
increasing the bias voltage to breakdown for a short time, the detector 
responsivity quickly relaxes to a stable level.  

\begin{figure}
  \begin{center}
   \FigureFile(80mm,65mm){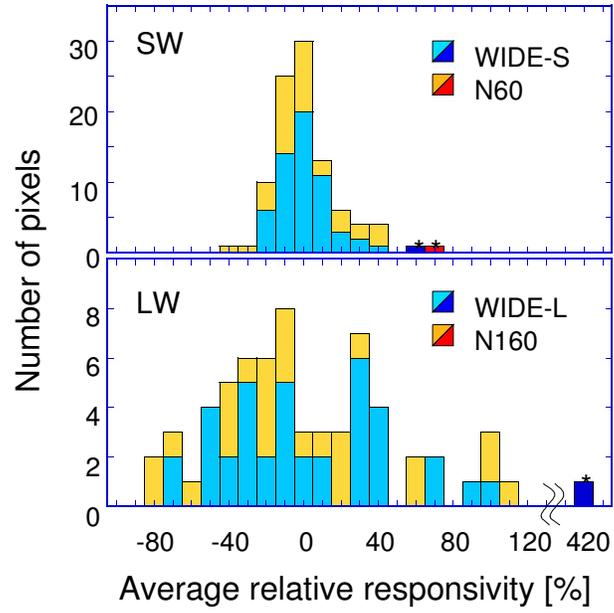}
  \end{center} 
   \caption{The histogram of the pixel variation for relative 
	responsivity.  The horizontal axis is normalized by the mean value 
	for each array excluding the anomalously high responsivity pixels, 
	which are marked by asterisks in the figure.  The variation in the 
	LW detector is larger than in the SW detector.
   }
  \label{fig:FIS_flat_field} 
\end{figure}

Finally, the pixel-to-pixel variations of the detector responsivity are 
shown in Fig.\ref{fig:FIS_flat_field}, which are the relative detector 
signals in the observation of flat sky. The unevenness of the detector 
responsivity could come from the non-uniformity of the effective detector 
bias, due to the offset of the readout electronics.  This is particularly 
an issue for the LW detector, because of its small bias voltage.  In 
addition, due to variations of the spectral response for each pixel in 
the LW detector, it has poor flatness as  compared to the SW detector.

\subsection{Photometric Calibration}

\begin{figure*}
  \begin{center}
   \FigureFile(150mm,90mm){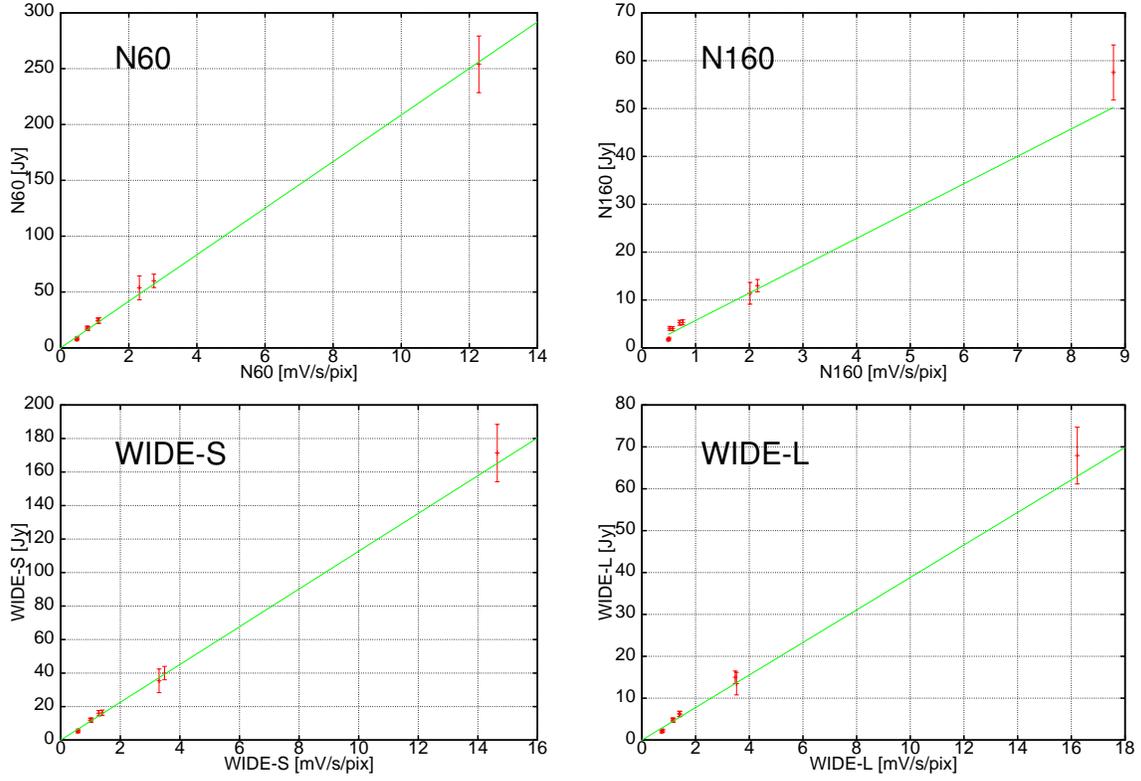}
  \end{center}
   \caption{The relation between the output signal and the model flux of 
	point sources for each band, observed in the pointing mode by FIS01. 
	The output signal is the integrated power within the definite 
	aperture size, and is corrected by the calibration lamps. Clear 
	linear relations between the source flux and the output signal are 
	seen in all bands for more than one order of magnitude in flux range.
   }
  \label{fig:abs_flux_cal} 
\end{figure*}

For absolute calibration, several kinds of astronomical sources are used 
--- well-modeled objects such as asteroids, planets, stars and galaxies, 
as well as the spectra of bright IR cirrus or interplanetary dust 
emission.  To calibrate the absolute flux from point like sources, the 
aperture photometry procedure must be defined, and then applied to the 
observations of asteroids, stars and galaxies with a wide range of 
fluxes.  The relation between the signal and the source flux for several 
sources has a good linear correlation as shown in 
Fig.\ref{fig:abs_flux_cal}.  The uncertainties  of the current signal to 
flux calibration are no more than 20\% for the N60 and WIDE-S bands, and 
30\% and 40\% for the WIDE-L and N160, respectively.  Our goal is to reach 
an absolute calibration accuracy of about 10\% in all bands.  We will 
achieve this by mitigation of various image artifacts described in this 
paper, and through analyses of many repeated observations of our network 
of well-known calibration sources, that are performed continually.

Comparing the signals of the sky brightness measured by FIS and DIRBE on 
COBE, which provide a well calibrated infrared sky map, we can obtain the 
absolute calibration for sky brightness.  
The observations of bright IR 
cirrus regions with no significant small scale structure are compared with 
values measured by DIRBE to make an absolute calibration.  The resulting 
calibration, however, disagrees with the calibration derived from the 
point like sources by a factor of about two.  The absolute calibration for 
diffuse sources still has a large uncertainty due to the difficulty of 
baseline estimation. Furthermore, the contribution of detector transient 
response differs between the two calibration methods.

According to the absolute calibration derived from the compact sources, 
the nominal detection limits evaluated from the signal-to-noise ratio of 
the detected sources are listed in Table \ref{tab:performance}.  The 
performance of the pointed observation is demonstrated by 
\citet{Matsuura07}.

\subsection{Spectroscopic Performance}

All the spectrometer functions work as they did in the laboratory.  After 
tuning the control sequence of the movable mirror, interferograms of the 
internal blackbody source and sky were measured.  The Fourier transformed 
spectra of the internal blackbody source are consistent with that taken in 
the laboratory, within a 10\% error after scaling the responsivity, which 
means that the laboratory optical performance is reproduced in space.  

\begin{figure}
  \begin{center}
   \FigureFile(80mm,70mm){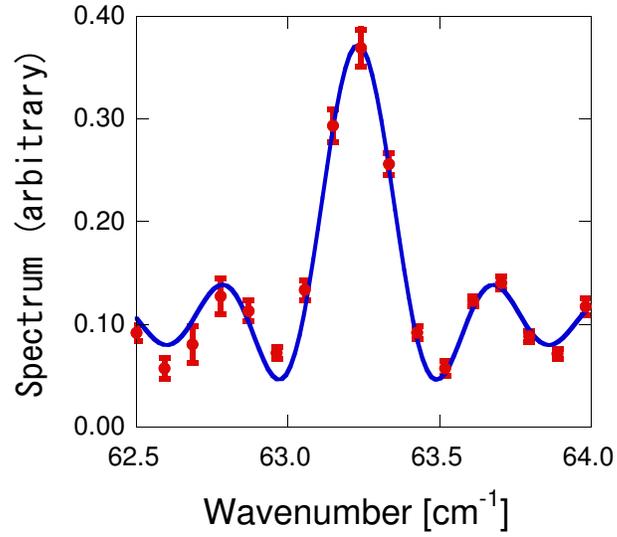}
  \end{center}
   \caption{A line shape detected by the FTS. This plot is a close up 
	around the [CII](157.74 $\micron$) line observed in M82.  Filled 
	circles with error bars are the measured spectrum without 
	apodization in arbitrary units.  The line is the best fit curve 
	using a sinc function. The derived spectral resolution from the 
	fitting is about 0.19 cm$^{-1}$, which is consistent with the 
	expected resolution of 0.185 cm$^{-1}$. 
   }
  \label{fig:FTS_line} 
\end{figure}

The data reduction to reproduce source spectra is difficult because the 
interferogram is distorted by the detector transient response as described 
above. Furthermore, a channel fringe in the interferogram also causes 
complications.  Through observations of well-known bright sources, line 
sensitivity and reproducibility of spectra were evaluated. The FTS system 
performance is almost same as the estimates from the laboratory 
measurements.  The line spectrum observed by the FTS is shown in the 
Fig.\ref{fig:FTS_line} as an example. The spectral resolution of the FTS 
in full resolution mode is about 0.19 cm$^{-1}$, which agrees well with 
the expected value of 0.185 cm$^{-1}$.  Interim detection limits of the 
FTS derived for on-source pixels from observations of bright sources are 
roughly 20 Jy, 50 Jy and 100 Jy for continuum spectra in wavenumber of 
$65 - 85$ cm$^{-1}$, $90 - 120$ cm$^{-1}$ and $> 120$ cm$^{-1}$, 
respectively, and, $3 \times 10^{-15}$ $W m^{-2}$ and 
$5 \times 10^{-14}$ $W m^{-2}$ for line emissions of 
[CII](158 $\micron$) and [OIII](88 $\micron$), respectively, which are 
$5 \sigma$ values for one-pointed observation.

\subsection{Comparison with Other Instruments}

FIS strives to provide an improved version of the all-sky survey performed 
by IRAS more than two decades ago.  The spectral coverage of FIS is 
extended to longer wavelength by the WIDE-L and N160 bands, which cover up 
to 180 $\micron$.  The longer wavelength coverage allows determination of 
the contribution of cold dust components, which perform important roles in 
the interaction of the interstellar medium and radiation field.  The 
advantage of longer wavelength capability has been demonstrated by the 
ISOPHOT Serendipity Survey (ISOSS) of ISO \citep{Stickel07b}. Although the 
sky coverage of ISOSS is about $15\%$ of the whole sky, the ISOSS 170 
$\micron$ Sky Atlas is utilized in variety of fields, especially, related 
to galaxies and cold galactic sources.  This precursor survey indicates 
the benefit of the longer wavelength bands in the FIS all-sky survey.  The 
high spatial resolution of the FIS all-sky survey is a great advantage for 
source detection and detailed mapping.  As shown in above, the spatial 
resolution of the FIS, which is about 0.7 acrmin for the N60 and WIDE-S 
bands, and about 1 arcmin for the WIDE-L and N160 bands, is more than 
five times better than IRAS even in the longer wavelength bands. The 
higher spatial resolution comes from the progress of detector technology, 
although the 60 cm diameter telescope 
of IRAS is comparable with the AKARI telescope \citep{Kaneda07a}.  These 
advantages of the FIS all-sky survey are demonstrated by \citet{Jeong07} and 
\citet{Doi07}.  

The point source flux levels at signal-to-noise ratio of five for 
one scan are listed in Table \ref{tab:performance}. We processed the observed data 
using the preliminary version of the data-processing pipeline for 
the all-sky survey data. We estimated the system sensitivity based on 
a series of observations of asteroids as calibration sources in 
the all-sky survey. We used Sussextractor, which is a point source 
extraction and photometry software dedicated for the AKARI all-sky 
survey \citep{Savage07}, to make the photometry of the asteroids.  
We derived noise levels by observing dark areas of the sky.  The 
estimated flux levels are significantly degraded from the detection 
limits estimated prior to the launch (see Table 3 in \cite{Matsuhara06}). 
We found several causes for the degradation. Firstly, we observed 
several types of excess noise in orbit. The current version of the 
pipeline successfully removed the effects of some types of noises.  
Secondly, we also observe the response to point sources is smaller 
than we expected. This is partly due to the PSFs described above, 
and partly due to the detector AC response.  Thirdly, we reduced the 
bias voltage for Wide-L and N160 in orbit to stabilize the behaviour 
of detectors after irradiation of high-energy particles.   
Furthermore, the actual source detection is degraded by various 
effects, e.g. frequent glitches due to high energy particles and 
low-frequency baseline fluctuation due to the change of the detector 
responsivity, in the data reduction process.   The potential performance of 
90 $\micron$ band (WIDE-S), however, is higher than that of the IRAS 
100$\micron$ band.  Unfortunately, the detection limits of the LW 
detector, which provides new wavelength bands, are also degraded.  
The potential performance of the longer wavelength bands is comparable to or 
better than the ISOSS.  The galaxy list of ISOSS has a 170 $\micron$ 
completeness limit of about 2 Jy \citep{Stickel07a}, and is useful 
for a cross calibration of the all-sky survey.  The performance of 
the all-sky survey evaluated from an initial mini-survey will be 
discussed by \citet{Shibai07b}.

The performance of the detailed observations using the pointing 
observation mode should be compared with the recent instrument MIPS on 
SST.  MIPS has advantages in spatial resolution and sensitivity, as a 
result of its larger telescope, small pixel scales, and long exposure 
capability.  The advantage of FIS is that it has four photometric bands 
between 50 and 180 $\micron$, whereas MIPS has only two bands (70 
$\micron$ and 160 $\micron$) in the corresponding wavelength range.  
Multi-band photometry with FIS is effective for determining a spectral 
energy distribution \citep{Kaneda07b, Suzuki07}.  The performance of 
the slow scan observation is demonstrated by \citet{Matsuura07}, who 
reported the detection limit at 90 $\micron$ (WIDE-S) achieves 26 mJy 
(3$\sigma$) using the observations at the Lockman Hole.  In the paper, 
they discuss about the source counts, and point out that the number of 
sources detected at 90 $\micron$ is significantly smaller at the faint 
end compared to the expected values from the model, which explains the 
MIPS source counts well.  This implies that FIS is a complementary 
instrument to MIPS in the SED coverage.  In the main observation 
phase, the low cirrus region near the South Ecliptic Pole (SEP) is 
observed by FIS intensively with almost the same sensitivity at the Lockman 
Hole for about 10 square degrees \citep{Matsuhara06}. The deep survey 
near the SEP will become a legacy survey of FIS for extragalactic studies.

The spectroscopic capability in the far-infrared region is a unique 
feature of FIS in contrast to MIPS.  Previously, LWS on ISO was available 
for far-infrared spectroscopy.  The wavelength coverage and the spectral 
resolution of LWS in the grating mode are comparable with the FIS 
spectrometer.  The sensitivity of FIS in the spectrometer mode is not so 
excellent as mention above. However, the spectrometer of FIS has the 
advantage of high observational efficiency.  Since it is an imaging FTS, FIS can 
take spectra with arcminute spatial resolution.  
For example, FIS could map the M82 galaxy with spectra in several pointed 
observations, which correspond to about one hour exposure time.  
Furthermore, the spectroscopic serendipity survey by parallel observations 
is expected to provide a unique dataset.

\section{Summary}

The Far-Infrared Surveyor was designed to survey the far-infrared region 
with four photometric bands within 50 -- 180 $\micron$, with high spatial 
resolution and sensitivity.   Additionally, a spectroscopic capability was 
installed as an imaging Fourier transform spectrometer.  All functions of 
FIS work very well in orbit.  FIS performance is demonstrated in the 
initial papers of this volume.  The all-sky survey is performed 
continuously and should provide a new generation all-sky catalog in the 
far-infrared.  In addition to the all-sky survey, a large area deep survey 
($\sim 10$ square degrees) near the South Ecliptic Pole and many scientific 
programs are being executed. 

To bring out the potential of the FIS instrument, the data processing 
methods are continuously being improved.   Since calibration data are 
accumulated constantly by the end of the mission life, the FIS data 
quality should be substantially better than that listed here.

\bigskip     
 
The AKARI project, previously named ASTRO-F, is managed and operated by 
the Institute of Space and Astronautical Science (ISAS) of Japan Aerospace 
Exploration Agency (JAXA) in collaboration with the groups in universities 
and research institutes in Japan, the European Space Agency, and Korean 
group. We thank all the members of the AKARI/ASTRO-F project for their 
continuous help and support.  
FIS was developed in collaboration with ISAS, Nagoya University, University 
of Tokyo, National Institute of Information and Communications Technology 
(NICT), National Astronomical Observatory of Japan (NAOJ), and other research 
institutes.  The AKARI/FIS All-Sky Survey data are processed by the 
international team which consists of members from the IOSG (Imprerial 
College, UK, Open University, UK,  University of Sussex, UK, and 
University of Groningen, Netherlands) Consortium, Seoul National University, 
Korea, and the Japanese AKARI team. The pointing reconstruction for the 
All-Sky Survey mode is performed by the pointing reconstruction team 
at European Space Astronomy Center (ESAC).  We thank all the members related 
to FIS for their intensive efforts toward creating new frontier.
M. Cohen's contribution to this paper was partially supported by a grant
from the American Astronomical Society.  We would like to express thanks to 
Dr. Raphael Moreno for providing flux models of giant planets.


\newpage

\begin{table*}
 \begin{center}
 \caption{Parameters of FIS AOTs}
 \label{tab:aot_list}
  \begin{tabular}{lccc}
   \hline \hline
   \multicolumn{1}{c}{{\bf AOT}} & {\bf FIS01} & {\bf FIS02} & {\bf FIS03} \\
   \hline
   observation mode & \multicolumn{2}{c}{slow scan} & pointing \\
   observation target & compact source & area mapping & spectroscopy \\ 
   \hline
   \multicolumn{4}{l}{{\bf parameters:}}\\
   - target position & \multicolumn{2}{c}{center of scan area} & source\\
   - reset interval\footnotemark[$*$] & \multicolumn{2}{c}{[CDS, 0.25 s, 0.5 s, 1.0 s, 2.0 s]} & [0.1 s, 0.25 s, 0.5 s, 1.0 s, (2.0 s)] \\
   - parameter 1\footnotemark[$*$] & \multicolumn{2}{c}{scan speed} & spectral resolution \\
   & \multicolumn{2}{c}{[8"/s, 15"/s, (30"/s)]} & [full res., SED] \\
   - parameter 2 & shift size & - & array selection\\
   & [70", 240"] & - & [MOD, SW, LW] \\
   \hline
   amount of data\footnotemark[$\dagger$] & \multicolumn{2}{c}{17.5 MB} & 49.8 MB \\
   \hline
   \\
   \multicolumn{4}{l}{\hbox to 0pt{\parbox{160mm}{\footnotesize
	Notes.
	\par\noindent
	\footnotemark[$*$] Parenthetic values of parameters are option.
	\par\noindent
	\footnotemark[$\dagger$] Data size that FIS generates in one pointing observation.
    }\hss}}

  \end{tabular} 
 \end{center}
\end{table*} 

\begin{table*}
 \begin{center}
 \caption{Flight Performance of FIS}
 \label{tab:performance}
  \begin{tabular}{lccccl}
   \hline \hline
   \multicolumn{1}{c}{{\bf BAND}} & {\bf N60} & {\bf WIDE-S} & {\bf WIDE-L} & {\bf N160} &\\
   \hline
   band center & $65$ & $90$ & $140$ & $160$ & [$\micron$]\\ 
   effective band width\footnotemark[$*$] & $21.7$ & $37.9$ & $52.4$ & $34.1$ & [$\micron$]\\ 
   \hline
   pixel scale & 26.8 & 26.8 & 44.2 & 44.2 & [arcsec]\\
   pixel pitch & 29.5 & 29.5 & 49.1 & 49.1 & [arcsec]\\
   \hline
   \multicolumn{5}{l}{{\bf point spread functions:}} \\
   - measured FWHM\footnotemark[$\dagger$] & 37 $\pm$ 1 & 39 $\pm$ 1 & 58 $\pm$ 3 & 61 $\pm$ 4 & [arcsec]\\
   \hline
   \multicolumn{5}{l}{{\bf flat field:}} \\ 
   - variation\footnotemark[$\ddagger$] & 19\% & 14\% & 43\% & 53\% &\\ 
   \hline
   \multicolumn{5}{l}{{\bf $5\sigma$ flux level:}} \\ 
   - survey mode\footnotemark[$\$$] & 2.4 & 0.55 & 1.4 & 6.3 & [Jy]\\ 
   - pointing mode\footnotemark[$\|$] & 110 & 34 & 350 & 1350 & [mJy] \\
   \hline
   \\
   \multicolumn{4}{l}{\hbox to 0pt{\parbox{120mm}{\footnotesize
	Notes.
	\par\noindent
	\footnotemark[$*$] Effective band width for the given band center under 
		the assumption of $\nu F_{\nu} = const$.
	\par\noindent
	\footnotemark[$\dagger$] FWHM derived from Gaussian fitting of PSFs constructed 
		from observations of asteroids in the photometer mode.
	\par\noindent
	\footnotemark[$\ddagger$] Relative variations (standard deviations) to the averages for each array. 
	\par\noindent
	\footnotemark[$\$$] Point source flux level with signal-to-noise ratio of 5 for one scan in the 
		all-sky survey mode.
	\par\noindent
	\footnotemark[$\|$] Point source flux level for one-pointed observation by FIS01 
		with 8"/sec scan speed and 2 sec integration time, which include both detector 
		and source noises.
    }\hss}}
  \end{tabular} 
 \end{center}
\end{table*} 

\begin{thebibliography}{}
\bibitem[Doi et al.(2002)]{Doi02} 
	Doi Y., et al., 2002, Adv. in Space Research, 30, 2099 
\bibitem[Doi et al.(2007)]{Doi07} 
	Doi Y., et al., 2007, \pasj, 00, 000
\bibitem[Fujiwara et al.(2003)]{Fujiwara03}
	Fujiwara M., Hirao T., Kawada M., Shibai H., Matsuura S., Kaneda H., Patrashin M.A., \& Nakagawa T.,
	2003, \ao, 42, 2166
\bibitem[Jeong et al.(2007)]{Jeong07}
	Jeong W-S., et al., 2007, \pasj, 00, 000
\bibitem[Kaneda et al.(2007a)]{Kaneda07a} 
	Kaneda H., Kim W., Onaka T., Wada T. Ira Y., Sakon I., \& Takagi T., 2007a, \pasj, 00, 000 
\bibitem[Kaneda et al.(2007b)]{Kaneda07b} 
	Kaneda H., et al., 2007b, \pasj, 00, 000 
\bibitem[Kessler et al.(1996)]{Kessler96}
	Kessler M.F., et al. 1996, \aap, 315, L27
\bibitem[Martin and Puplett(1969)]{Martin69}
	Martin D.H., \& Puplett E., 1969, Infrared Phys., 10, 105 
\bibitem[Matsuhara et al.(2006)]{Matsuhara06}
	Matsuhara, H., et al., 2006, \pasj, 58, 673
\bibitem[Matsuura et al.(2007)]{Matsuura07}
	Matsuura, S., et al., 2007, \pasj, 00, 000
\bibitem[Murakami et al.(2007)]{Murakami07}
	Murakami H., et al. 2007, \pasj, 00, 000
\bibitem[Nagata et al.(2004)]{Nagata04} 
	Nagata H., Shibai H., Hirao T., Watabe T., Noda M., Hibi Y., Kawada M., \& Nakagawa T.,
	2004, IEEE Trans. on Elec. Devices, 51, 270 
\bibitem[Nakagawa et al.(2007)]{Nakagawa07}
	Nakagawa T., et al., 2007, \pasj, 00, 0000
\bibitem[Onaka et al.(2007)]{Onaka07}
	Onaka T., et al. 2007, \pasj, 00, 000
\bibitem[Savage et al. (2007)]{Savage07}
	Savage R., et al. 2007 in preparation
\bibitem[Shibai H.(2007a)]{Shibai07}
	Shibai H., 2007a, Adv. in Space Research, accepted for publication
\bibitem[Shibai et al.(2007b)]{Shibai07b}
	Shibai H., 2007b, in preparation
\bibitem[Stickel et al.(2007a)]{Stickel07a}
	Stickel, M., Klaas, U., and Lemke, D., 2007a, \aap, 466, 831
\bibitem[Stickel et al.(2007b)]{Stickel07b}
	Stickel, M., Krause, O., Klaas, U., and Lemke, D., 2007b, \aap, 466, 1205
\bibitem[Suzuki et al.(2007)]{Suzuki07}
	Suzuki T., Kaneda H., Nakagawa T., Makiuti S., Doi Y., \& Shibai H., 2007, \pasj, 00, 000
\bibitem[Werner et al.(2004)]{Werner04}
	Werner M.W., et al. 2004, \apjs, 154, 1
\end{thebibliography}
\end{document}